\documentclass[journal,hideappendix]{vgtc}        % final (journal style) without appendices
\onlineid{0}

\vgtccategory{Research}

\vgtcpapertype{please specify}

\title{Towards an Understanding and Explanation for Mixed-Initiative Artificial Scientific Text Detection}

\author{%
  Luoxuan Weng,
  Minfeng Zhu, 
  Kam~Kwai~Wong, 
  Shi Liu,
  Jiashun Sun,
  Hang Zhu,
  Dongming Han, and 
  Wei Chen
}

\authorfooter{
  \item
    Luoxuan Weng, Minfeng Zhu, Shi Liu, Jiashun Sun, Hang Zhu, Dongming Han, and Wei Chen are with the State Key Lab of CAD\&CG, Zhejiang University. E-mail: \{lukeweng, minfeng\_zhu, zju\_ls, sunjs, 3190102785, dongminghan, chenvis\}@zju.edu.cn.
  \item
    KK Wong is with the Hong Kong University of Science and Technology. E-mail: kkwongar@connect.ust.hk.
}

\abstract{
Large language models (LLMs) have gained popularity in various fields for their exceptional capability of generating human-like text. Their potential misuse has raised social concerns about plagiarism in academic contexts. However, effective artificial scientific text detection is a non-trivial task due to several challenges, including 1) the lack of a clear understanding of the differences between machine-generated and human-written scientific text, 2) the poor generalization performance of existing methods caused by out-of-distribution issues, and 3) the limited support for human-machine collaboration with sufficient interpretability during the detection process.
In this paper, we first identify the critical distinctions between machine-generated and human-written scientific text through a quantitative experiment. Then, we propose a mixed-initiative workflow that combines human experts' prior knowledge with machine intelligence, along with a visual analytics prototype to facilitate efficient and trustworthy scientific text detection. Finally, we demonstrate the effectiveness of our approach through two case studies and a controlled user study with proficient researchers. We also provide design implications for interactive artificial text detection tools in high-stakes decision-making scenarios.

}


\teaser{
  \centering
  \includegraphics[width=\linewidth]{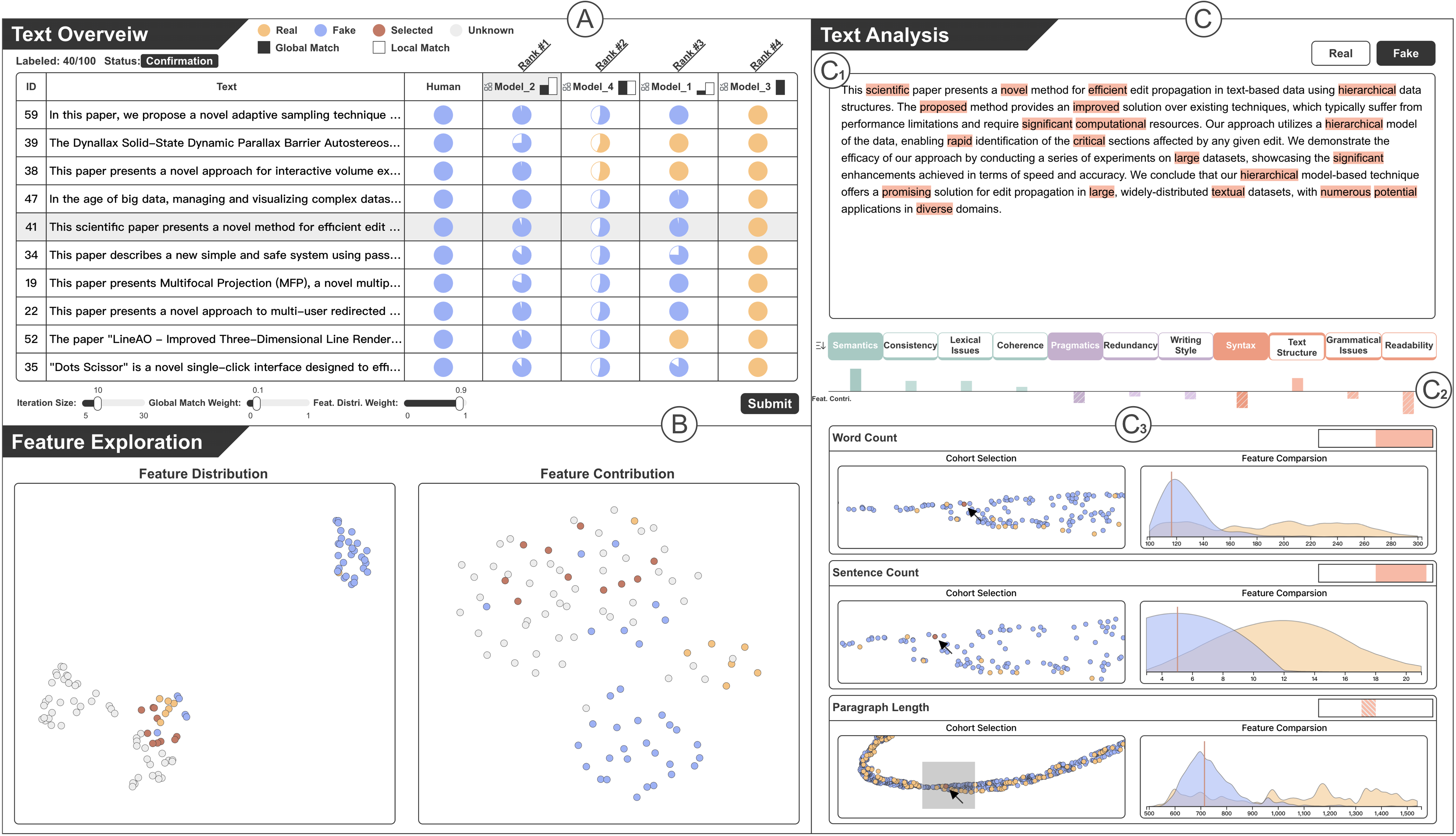}
  \caption{We present a prototype system for detecting artificial scientific text. (A) The \emph{Text Overview} panel allows users to provide weak annotations for text excerpts based on their prior knowledge and compare them with multiple ML models' decisions. (B) The \emph{Feature Exploration} panel presents an overview of the feature distribution and contribution of all text excerpts to be detected. (C) The \emph{Text Analysis} panel offers contribution-level, distribution-level, and excerpt-level feature analysis to facilitate an effective and interpretable detection process. By iteratively interacting with the three coordinated panels, users can perform effective and efficient detection of multiple scientific text through a mixed-initiative workflow.}
  \label{fig:teaser}
}

\graphicspath{{figs/}{figures/}{pictures/}{images/}{./}} % where to search for the images

\usepackage{tabu}                      % only used for the table example
\usepackage{booktabs}                  % only used for the table example
\usepackage{lipsum}                    % used to generate placeholder text
\usepackage{mwe}                       % used to generate placeholder figures

\usepackage{mathptmx}                  % use matching math font

\usepackage{url}

\usepackage{multicol}
\usepackage{rotating}
\usepackage{multirow}
\newcommand{\rot}[1]{\rotatebox[origin=l]{45}{#1}}

\usepackage{xspace,xpunctuate}
\usepackage{tabularx}

\newcommand{\ie}{\textit{i.e.},\xspace}
\newcommand{\etal}{\xspace\textit{et al.}\xspace}
\newcommand{\eg}{\textit{e.g.},\xspace}
\newcommand{\icon}[1]{\includegraphics[height=\fontcharht\font`\B]{#1}}

\begin{document}

\firstsection{Introduction}
\maketitle

The recent emergence of large language models (LLMs) has significantly enhanced the diversity, control, and quality of machine-generated text\cite{crothers2022machine}. For instance, ChatGPT\cite{chatgpt} gained tremendous public attention for its ability to generate plausible, coherent, and human-like text in a conversational manner (or \emph{prompting}\cite{9908590}). Nonetheless, the widespread accessibility of powerful LLMs has raised concerns regarding their potential misuse, including the propagation of fake news\cite{zellers2019defending, pagnoni2022threat}, fraudulent online reviews\cite{kowalczyk2022detecting}, and social media spam\cite{MIRSKY2023103006}. These concerns are particularly acute in academic and educational contexts\cite{ma2023abstract, rodriguez2022cross}. Academic conferences and journals such as ACL and ICML have established strict guidelines for the cautious and responsible use of LLMs\cite{acl, icml}. In this paper, we focus on the detection of scientific text, as there is an urgent need to prevent cheating and maintain academic integrity.

Recently, much attention has been devoted to artificial text detection in industry and academia. The growing demand for detecting machine-generated text has led to the development of several online tools such as GPTZero\cite{gptzero}. Meanwhile, academic research has also proposed numerous detection methods, including feature-based\cite{frohling2021feature, nguyen2017identifying} and transformer-based\cite{rodriguez2022cross, mitrovic2023chatgpt} models. Although existing methods can achieve high accuracy in specific domains such as fake news discrimination\cite{zellers2019defending}, they are less effective when applied to detect machine-generated scientific text due to the following reasons.

Firstly, the critical distinctions between machine-generated and human-written scientific text remain underexplored. Previous research findings\cite{guo2023close, dou2022gpt, ma2023abstract} are either not specific to scientific text or lack comprehensive and quantitative user studies. Understanding these differences is crucial for designing effective detection algorithms and enhancing the ability of human experts to identify suspect manuscripts.

Secondly, the performance of existing methods often lacks generalizability due to out-of-distribution (OOD) issues. Numerous LLMs can generate scientific text with varying feature distributions that are difficult to distinguish using a single detector, caused by diverse sampling methods of LLMs\cite{frohling2021feature} or cross-domain adaptation issues\cite{rodriguez2022cross}. Moreover, different detectors might produce incorrect and conflicting results when confronted with multiple scientific text of unknown origins.

Thirdly, the incorporation of human agency is limited. Recent studies\cite{ippolito2020automatic,ma2023abstract} have found that expert human reviewers outperform automated detectors in identifying certain shortcomings of machine-generated scientific text. This suggests that integrating human experts' prior knowledge is promising for improving the fairness, interpretability, and reliability of artificial text detection, as argued in \cite{ippolito2020automatic, crothers2022machine, jawahar2020automatic}. However, current approaches such as GLTR\cite{gehrmann2019gltr} lack sufficient support for human-machine collaboration and interpretation of machine learning (ML) models' decisions. In high-stakes decision-making scenarios like scientific text detection, solely relying on models' decisions reduces the reliability and trustworthiness of the detection process.

To address the aforementioned challenges, we first conduct a formative study with a group of 12 experts to understand the critical distinctions and corresponding statistical features between machine-generated and human-written scientific text, and formulate design requirements through user interviews. Our results show that, while knowledgeable human experts can holistically identify machine-generated scientific text, they seek evidence from ML models to support their decisions. Therefore, we propose a mixed-initiative workflow that combines human experts' prior knowledge with ML models trained using the identified features.
% The workflow 1) presents text excerpts based on feature similarities and ranks multiple ML models during each iteration to improve the effectiveness and efficiency, and 2) leverages explainable artificial intelligence (XAI) techniques to provide contribution-based explanations for ML models' decisions to enhance the interpretability and reliability of the detection process.
Our workflow presents text excerpts based on feature similarities and ranks multiple ML models during each iteration to mitigate the OOD issue when handling multi-sourced text. It also provides multiple levels of feature analysis leveraging explainable artificial intelligence (XAI) techniques to enhance the interpretability and reliability of the human-machine detection process.
Accordingly, we design and implement a visual analytics prototype system. Case studies and a controlled user study demonstrate the capability of our approach in facilitating artificial scientific text detection.

In summary, the major contributions of this work include:

\begin{itemize}
\vspace{-1.5mm}
\item
A formative study that identifies critical distinctions and summarizes design requirements for artificial scientific text detection.
    
\vspace{-1.5mm}
\item
A novel workflow and a prototype system that combines human intelligence with machine learning techniques to facilitate artificial text detection.

\vspace{-1.5mm}
\item
Two case studies and a controlled user study to demonstrate the effectiveness of our proposed approach.

\end{itemize}
\section{Related Work}
\label{sec:related_work}

\subsection{Artificial Text Detection}
Various techniques have been proposed to differentiate between machine-generated and human-written text. Crothers\etal\cite{crothers2022machine} presented a comprehensive review of automatic detection of machine-generated text. Most existing work can be broadly categorized into two groups: feature-based methods and deep-learning-based methods.

Feature-based methods\cite{crothers2022adversarial, frohling2021feature, guo2023close, nguyen2017identifying, kowalczyk2022detecting} utilize statistical distinctions between machine-generated and human-written text to train ML models for classification. These statistical features include basic features (\eg word and sentence length\cite{ma2023abstract, frohling2021feature}), frequency features (\eg TF-IDF\cite{solaiman2019release}), fluency features (\eg Gunning-Fog Index and Flesch Index\cite{crothers2022adversarial}), \emph{etc.} Perplexity, another metric used to measure the efficacy of a model in predicting the next word in a sequence, is also commonly adopted to identify machine-generated text\cite{ma2023abstract, guo2023close}. However, feature-based methods are hindered by different sampling methods or model sizes\cite{holtzman2019curious, zellers2019defending}, whereas our proposed approach incorporates multiple models with human expertise to address these limitations.

Deep-learning-based methods\cite{mitrovic2023chatgpt, rodriguez2022cross, pu2022deepfake, stiff2022detecting, pagnoni2022threat, ippolito2020automatic} utilize neural networks or language models to differentiate between machine-generated and human-written text.
% commonly involve the fine-tuning of pre-trained language models such as BERT\cite{devlin2018bert} and RoBERTa\cite{liu2019roberta} on paired datasets of machine-generated and human-written text for classification.
For instance, Zellers\etal\cite{zellers2019defending} proposed the Grover model for generating and detecting fake news, demonstrating the potential of generative models in discrimination, which has also been observed in other studies\cite{radford2019language, solaiman2019release}. Fine-tuning approaches, particularly those utilizing large bi-directional language models like RoBERTa, still represent the state-of-the-art for artificial text detection\cite{solaiman2019release, crothers2022machine}. While cross-domain adaptation has shown significant improvement with a few hundred out-of-domain samples\cite{rodriguez2022cross}, collecting balanced training data for general-purpose detection models in real-life scenarios remains a significant challenge\cite{jawahar2020automatic, crothers2022machine, bakhtin2019real}. Moreover, the black-box nature of most deep learning models impedes their adoption in high-stakes decision-making scenarios like scientific text detection.

% Other studies have also utilized watermarking\cite{kirchenbauer2023watermark}, fact verification\cite{zhong2020neural}, attention map topology\cite{kushnareva2021artificial}, and probability curvature\cite{mitchell2023detectgpt} to perform the detection.
% Additionally, empirical research\cite{dugan2023real, clark2021all, dou2022gpt, ippolito2020automatic} has investigated humans' ability in identifying machine-generated text. Notably, Dou\etal\cite{dou2022gpt} proposed the Scarecrow framework to identify 10 error categories commonly made by GPT-3 and demonstrated that human annotators outperformed a span classification model in detecting certain errors.
In addition to automatic detection approaches, empirical research\cite{dugan2023real, clark2021all, dou2022gpt, ippolito2020automatic} has investigated humans' ability of detecting machine-generated text. For instance, human experts achieve high precision in detecting certain errors like technical jargon\cite{dou2022gpt} and identifying fake scientific abstracts\cite{ma2023abstract}. Inspired by these findings and the concept of GLTR\cite{gehrmann2019gltr} which incorporates human agency to facilitate detection, we propose a mixed-initiative workflow that more effectively integrates human intelligence.

\subsection{Explainable Artificial Intelligence}

XAI techniques aim to enhance the interpretability and reliability of AI models, enabling humans to understand how these models make their decisions. Several surveys\cite{10.1145/3359786, electronics8080832, 8400040, 8466590} have been proposed to provide a comprehensive overview of XAI approaches, including intrinsic and post-hoc interpretability\cite{molnar2020interpretable}. Intrinsic interpretability\cite{EIRASFRANCO2019113141, article, 8612916} refers to ML models that can be understood and explained directly from their design and architecture, such as decision trees and generalized additive algorithms. Visualization techniques have also been proposed to facilitate intrinsic interpretability. For instance, Dingen\etal\cite{8464305} presented RegressionExplorer to interactively explore logistic regression models in clinical biostatistics. Similarly, Neto\etal\cite{9222255} proposed Explainable Matrix for interpreting complex random forest ensembles. On the other hand, post-hoc interpretability techniques\cite{9229232, NIPS2017_8a20a862, 10.1145/2939672.2939778, Ribeiro_Singh_Guestrin_2018} are applied after model training, specifically for black-box models like deep neural networks that are not intrinsically interpretable. Such methods are model-agnostic and can be utilized to explain the decisions of any ML models. One of the typical approaches is to train a surrogate model to approximate the outputs of the original black-box model\cite{molnar2020interpretable}.  For instance, LIME\cite{10.1145/2939672.2939778} trains local surrogate models to explain individual predictions. Another popular method, SHAP (SHapley Additive exPlanations)\cite{NIPS2017_8a20a862}, which is based on game theory\cite{shapley1953value}, has been widely applied in various XAI systems\cite{9555810, collaris2022strategyatlas, 9552849}. In this work, we leverage SHAP values to offer contribution-based explanations for models' decisions in scientific text detection.

\subsection{Mixed-Initiative Systems}

Initially introduced by Horvitz\cite{horvitz1999principles}, mixed-initiative systems aim to enhance collaboration between humans and machines in decision-making processes. In recent years, there has been a significant amount of work in the visualization field related to mixed-initiative visual analytics systems\cite{cook2015mixed, wall2017podium, felix2018exploratory, perez2022typology, wenskovitch2021beyond, crouser2016toward, husain2021mixed, cabrera2019fairvis, pister2020integrating}. These systems leverage both human and machine intelligence to explore and analyze complex data utilizing innovative visual designs. For instance, Wall\etal\cite{wall2017podium} proposed the Podium system that enabled users to rank multi-attribute data based on their holistic understanding and explore their subjective preferences through mixed-intiative visual analytics. Our work is also based on a similar assumption derived from our formative study, that knowledgeable human experts can only holistically identify machine-generated scientific text and require evidence from ML models to support their decisions. Additionally, Pister\etal\cite{pister2020integrating}'s work on integrating prior knowledge for social network clustering has motivated our proposed workflow. Unlike clustering, we focus on the binary classification task of detecting artificial text.

Our work involves a binary labeling process, which is similar to previous works on mixed-initiative labeling systems.
% Suh\etal\cite{suh2016label} statistically demonstrated the superiority of mixed-initiative classifier training in reducing label complexity
For instance, Felix\etal\cite{felix2018exploratory} proposed an interactive visual data analysis method to facilitate document labeling with machine recommendations. Similarly, Choi\etal\cite{choi2019aila} developed the Attentive Interactive Labeling Assistant to visually highlight words and improve the efficiency of document labeling. More recently, Alsaid\etal\cite{alsaid2022datascope} integrated dimensionality reduction to develop an interactive method for labeling video and image data more efficiently and effectively. Furthermore, the visualization community has explored visual-interactive labeling which combines VA with ML techniques\cite{bernard2017comparing, bernard2018vial, bernard2021taxonomy, sevastjanova2021questioncomb}. In contrast to prior works that primarily focused on enhancing the effectiveness and efficiency of the labeling process\cite{bernard2017comparing}, we extend this line of research by prioritizing the reliability of high-stakes decision-making scenarios like scientific text detection.

\section{Formative Study}
\label{sec:formative_study}

\begingroup
\renewcommand{\arraystretch}{1.3}
\begin{table*}[t]
\small
\centering
\caption{Summary of distinctions between machine-generated and human-written scientific text.}
\label{tab:distinctions}
\setlength\tabcolsep{3.5pt}{
\begin{tabular}{llll}
\toprule
\textbf{Dimension}  & \textbf{Subcategory}        & \textbf{Description}                                                                                 & \textbf{Features}                                             \\ \hline
\multirow{3}*{\rot{Syntax}}     & Grammatical Issues & The correctness and accuracy of using words, phrases and clauses in a sentence              & Part-of-Speech Tag Frequency, Punctuation Frequency  \\ \cline{2-4} 
           & Text Structure     & The organization and arrangement of sentences and paragraphs in a text                      & Paragraph Length, Word/Sentence Count, etc.          \\ \cline{2-4} 
           & Readability        & The ease of reading and understanding the text                                              & Gunning-Fog Index, Flesch Reading Ease               \\ \hline
\multirow{3}*{\rot{Semantics}}  & Lexical Issues     & The choice and usage of words that convey the intended meaning and tone of a text           & Google's Top Word Frequency, TF-IDF, etc.            \\ \cline{2-4} 
           & Consistency        & The agreement and harmony of words, phrases and sentences in a text                         & Average Cosine Similarity between Sentence and Title \\ \cline{2-4} 
           & Coherence          & The logical connection and relation between sentences and paragraphs in a text              & Average Cosine Similarity between Sentences    \\ \hline
\multirow{6}*{\rot{Pragmatics}} & Redundancy         & The unnecessary repetition of information in a text                                         & Unigram/Bigram/Trigram Overlap of Words/PoS Tags     \\ \cline{2-4} 
           & Writing Style      & The distinctive manner of expressing ideas, opinions or emotions in a text                  & SciBert\cite{beltagy-etal-2019-scibert} Embedding                                    \\ \cline{2-4} 
           & Self-Contradiction & The inconsistency or conflict between different parts or aspects of a text                  & Not Applicable                                                 \\ \cline{2-4} 
           & Commonsense        & The general knowledge or understanding that is expected from the reader/writer of a text & Not Applicable                                                 \\ \cline{2-4} 
           & Factuality         & The level of accurate and verifiable information in a text                                  & Not Applicable    \\ \cline{2-4}
           & Specificity        & The level of detail in a text to support the main points & Not Applicable      
           \\ \bottomrule
\end{tabular}
}
\end{table*}
\endgroup

\begin{figure*}[ht]
  \centering
  \includegraphics[width=\textwidth]{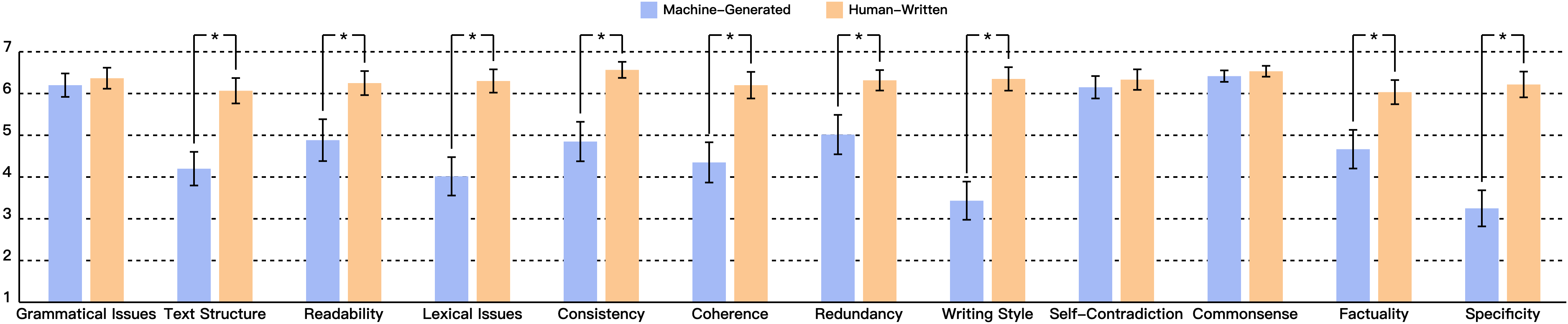}
  \caption{Average ratings of distinction categories on a 7-point Likert scale ($*:p<.05$), where error bars represent 95\% confidence intervals.}
  \label{fig:ratings}
  \vspace*{-3mm}
\end{figure*}

\begin{figure*}[ht]
  \centering
  \includegraphics[width=\textwidth]{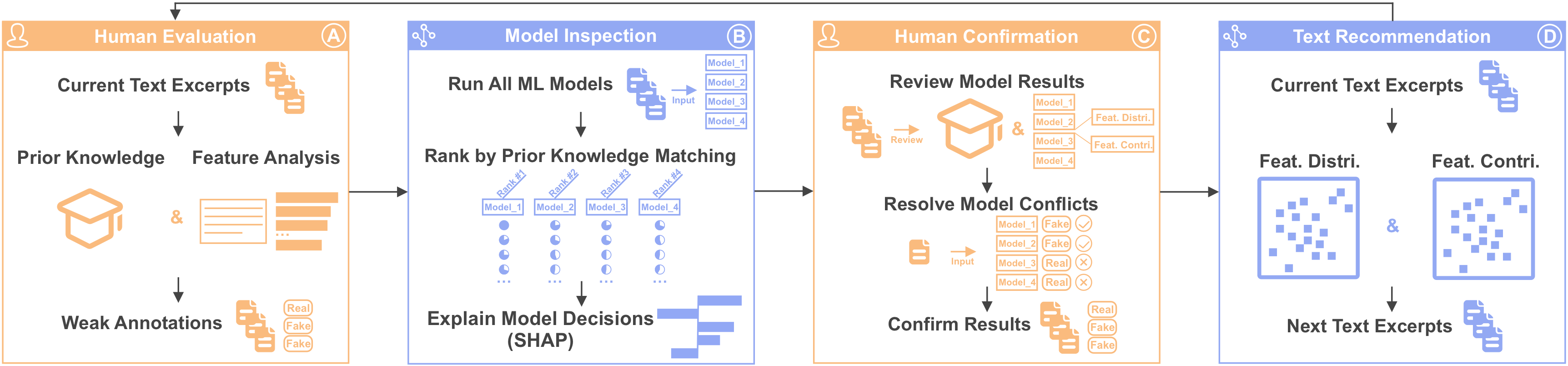}
  \caption{The proposed workflow comprises four iterative stages: (A) human experts provide weak annotations for current text excerpts based on their prior knowledge, (B) ML models are run and ranked based on their \emph{level of prior knowledge matching}, (C) human experts review and confirm results for current text excerpts based on the analysis of models' and their own decisions, and (D) next text excerpts are recommended based on their weighted similarities to current text excerpts.}
  \label{fig:workflow}
  \vspace*{-3mm}
\end{figure*}

Our workflow is tailored for proficient researchers with a sound comprehension of the subject matter. We collected paired datasets related to visualization research, and collaborated with visualization experts, including senior PhD students and editors of relevant journals. The formative study spanned 4 months, during which we conducted bi-weekly semi-structured interviews with the experts to comprehend their primary concerns regarding existing automatic detection methods for scientific text. Through quantitative experiments, we identified key statistical features that can distinguish machine-generated text from human-written text. We consolidated the design requirements and progressively refined them based on user feedback to guide the design and development of our proposed workflow and prototype system.

\subsection{Experiment Design}
\textbf{Data Collection.} 
We extracted the titles and original abstracts from the dataset provided by Narechania\etal\cite{9552447}. We then used the metadata to generate abstracts through the official APIs for ChatGPT and GPT-3 with the prompt \emph{‘suppose you are the author, write a short abstract for the scientific paper titled with TITLE’}. For GPT-2, we used the first sentence of each original abstract as a prompt. Therefore, we constructed three paired datasets containing machine-generated and human-written abstracts for the same title.

\textbf{Participants.} We recruited 12 researchers (9 males and 3 females) from a local university. Each participant has a minimum of 3 years' experience in conducting research on visualization, with 10 possessing previous knowledge of utilizing AI tools (\eg Grammarly\cite{grammarly}) for proofreading or paraphrasing.

\textbf{Settings.} Based on previous literature\cite{frohling2021feature, dou2022gpt, ma2023abstract} and discussions with experts, we identified three major dimensions of distinctions, \ie \emph{syntax}, \emph{semantics}, and \emph{pragmatics}\cite{gleason2022development}, each of which consists of several subcategories (Tab.~\ref{tab:distinctions}). While we acknowledge that we may not have exhaustively captured all possible distinctions, we believe that our derived categories offer valuable insights for future empirical research.
% Additionally, we fine-tuned a strong baseline model, OpenAI's RoBERTa-based detector\cite{solaiman2019release}, using our collected dataset. The detector's output can serve as a reference for participants to contemplate and interpret the current state-of-the-art in artificial text detection and assess its practical application.

\textbf{Procedure.} We first introduced the study settings and offered a comprehensive explanation of the meaning and coverage of each distinction category. Upon ensuring that the participants had familiarized themselves with the study, we sampled five pairs of machine-generated and human-written abstracts with the same title from our dataset without specifying their labels. We then asked the participants to rate the \emph{level of quality} for each distinction category of each abstract pair on a 7-point Likert scale from 1 (\emph{low quality}) to 7 (\emph{high quality}). Finally, we collected and analyzed their feedback and conducted a post-study interview to comprehend their evaluation criteria for machine-generated text. Notably, our study extended previous research\cite{ma2023abstract} by presenting a more fine-grained distinction framework and explicitly requesting participants to rate each distinction category separately.

\subsection{Results Analysis}
Figure~\ref{fig:ratings} illustrates the average ratings of various distinction categories used to evaluate machine-generated scientific text. To determine the most important features for humans to distinguish such text, we conducted pairwise comparisons of ratings of abstract pairs within each distinction category using a post-hoc Wilcoxon signed-rank test.
% Statistical significance is determined for $p<.05$($*$), $p<.01$($**$), $p<.001$($***$).
If the ratings of abstract pairs are significantly different, the corresponding distinction category should be considered a \emph{critical distinction}, as participants can accurately compare the quality of the scientific text in that category. Based on our analysis of the results and post-study interviews with the participants, we summarize our key findings below.
% Further information regarding the experiment results and analysis can be found in the supplementary material.

The \emph{Syntax} dimension refers to the rules and patterns of language that govern how words and sentences are formed and organized. Our results revealed that human-written scientific text achieved higher ratings in the \emph{Syntax} dimension compared to machine-generated text. Moreover, we observed significant differences in \emph{Text Structure} and \emph{Readability}, but not in \emph{Grammatical Issues}, suggesting that machine-generated scientific text generally exhibit grammatical correctness, but struggle with effective sentence and paragraph organization, leading to suboptimal communication of scientific ideas to readers. For instance, some participants complained that the machine-generated text were \emph{`very hard to follow sometimes'} (P3), which might be attributed to issues such as \emph{`excessively long sentences'} (P3), \emph{`inconsistent use of verb tense'} (P2), or \emph{`overuse of passive voice'} (P9). In contrast, human-written scientific text tend to have a more structured presentation of ideas, facilitating better understanding and interpretation of the content.

The \emph{Semantics} dimension refers to the meaning and interpretation of words and sentences in context. Notably, human-written text in this dimension received higher average ratings as well. Our results revealed significant differences in all three subcategories: \emph{Lexical Issues}, \emph{Consistency}, and \emph{Coherence}, indicating that machine-generated scientific text can be easily distinguished by human experts in terms of semantics. Almost half of the participants (5/12) reported inconsistencies between the title and abstract and incoherencies between sentences. For instance, P4 noted that \emph{`the title is related to time-varying data analysis while the entire abstract keeps talking about focal point extraction'}, while P11 observed that \emph{`most generated text were not logically coherent due to unexpected or unreasonable expressions'}. Additionally, for \emph{Lexical Issues}, some participants complained about \emph{`the lack of lexical diversity'} (P6, P8) and \emph{`informal use of certain words'} (P2) in machine-generated text. These findings confirmed previous literature\cite{ma2023abstract} that suggested machine-generated scientific text still struggled with semantic consistency and coherence when conveying complex ideas and insights. Interestingly, although most participants did not detect any misspellings or vocabulary mistakes, those who were native English speakers all agreed that machine-generated text sounded \emph{`more natural'} (P6) due to \emph{`a better choice of vocabulary'} (P7). This may be attributed to the use of reinforcement learning from human feedback (RLHF)\cite{christiano2017deep} in LLMs like ChatGPT during training, enabling them to sound more human-like. Future research is necessary to investigate this further.

The \emph{Pragmatics} dimension refers to the purpose and effect of language in communication and interaction. As expected, human-written text in this dimension received higher average ratings, particularly in terms of \emph{Writing Style} and \emph{Specificity}. Additionally, significant differences were observed in \emph{Redundancy}, \emph{Writing Style}, \emph{Factuality}, and \emph{Specificity}, but not in \emph{Self-Contradiction} and \emph{Commonsense}. Specifically, participants expressed dissatisfaction with machine-generated scientific text due to \emph{`inadequate details in background, motivation, method and evaluation'} (P2) and \emph{`low formality regarding the overall writing style'} (P7). However, while some of the scientific text were generally much detailed by using well-designed prompts, some participants noted that \emph{`longer and more detailed generated text were often less precise and coherent'} (P5). Our findings aligned with previous literature that machine-generated text were limited due to a "lack of purpose and functionality"\cite{frohling2021feature}. Nevertheless, participants did not report errors related to \emph{Self-Contradiction} and \emph{Commonsense}, contradicting findings in \cite{dou2022gpt}. We attribute this to two primary reasons. First, our datasets mostly comprise short scientific text (less than 250 words), which are less likely to contain self-contradictions compared to longer text. Second, scientific text generally contain less commonsense knowledge than fake news\cite{monti2019fake} or online reviews\cite{jawahar2020automatic}, resulting in a lower likelihood of commonsense errors.

In summary, we explored the critical distinctions that human experts rely on to differentiate between machine-generated and human-written scientific text in terms of \emph{syntax}, \emph{semantics}, and \emph{pragmatics}. Results revealed that participants rated human-written scientific text higher on all distinction categories, and significant differences were observed in 9 out of 12 categories, providing empirical evidence that a gap still exists between human-written and machine-generated text in scientific writing\cite{ma2023abstract}. The derived critical distinctions were then used to identify \emph{key statistical features} (Tab.~\ref{tab:distinctions}) informed by previous literature\cite{ma2023abstract, frohling2021feature} to train ML models for our proposed workflow (Sec. \ref{sec:workflow}).

\subsection{Design Requirements}
The experiment revealed that human experts could holistically detect machine-generated scientific text based on the quality of critical distinctions, but may not understand the statistical features contributing to their decisions. To this end, we aim to design a mixed-initiative workflow that incorporates both human and machine intelligence and supports interpretation of ML models to enhance the effectiveness and efficiency of the detection process. Throughout the formative study, we solicited input from the experts on their daily workflow for identifying scientific text, their perspectives on existing automatic detection methods, and their requirements for a detection process involving human intervention. Subsequently, we summarized the insights obtained to establish a comprehensive set of design requirements.

\textbf{R1: Support adequate human involvement.} In the context of scientific text detection, our experts have raised concerns about the limited or non-existent incorporation of human agency in existing artificial detection methods. This inadequacy leads to decreased efficacy, particularly in scenarios that demand domain expertise to ensure reliable and trustworthy decision-making. To address this issue, our experts suggested sufficient integration of their prior knowledge into the scientific text detection process, and highlighted the significance of interactive visualization techniques to facilitate effective human-machine collaboration. As such, our workflow should support sufficient human involvement through a user-initiated mixed-initiative procedure.

\textbf{R2: Integrate decisions from multiple ML models.} Although experts possessed the knowledge to differentiate between machine-generated and human-written scientific text, they recognized the potential value in using existing ML models to complement and augment their judgments. As numerous detection approaches have been proposed recently, our experts hoped to leverage multiple ML models to enhance their decision-making process, similar to ensemble learning techniques\cite{dong2020survey}. However, the varying performances of different models under different conditions present a challenge in selecting the most appropriate model for a specific use case. Therefore, our workflow should integrate multiple ML models and provide a mechanism for experts to select the most reliable model for their needs with ease.

\textbf{R3: Explain the decisions of ML models.} One of the major concerns raised by our experts is the lack of interpretability in most black-box detection models. In high-stakes decision-making scenarios like scientific text detection, editors and reviewers must thoroughly understand the reasons behind ML models' decisions to accept or reject the manuscript to ensure academic integrity and credibility. Furthermore, our experts pointed out that contribution-based explanations alone may be insufficient to fully comprehend models' decisions, especially when typical feature values are unknown. Accordingly, it is essential to employ XAI techniques along with cohort-level feature analysis to enhance the interpretability and reliability of our workflow, in line with the recommendations of our experts and previous research literature\cite{crothers2022machine}.

\textbf{R4: Analyze multiple text effectively and efficiently.} In the context of scientific text detection, experts are often presented with a large number of submitted papers (\eg 500) and are required to identify the potentially suspicious ones in a timely manner. However, existing industry products for artificial text detection (\eg GPTZero) only allow users to upload text one by one, which is time-consuming and labor-intensive. While it is possible to deploy an ML model offline and process multiple text simultaneously, our experts have found it challenging to quickly determine the most effective model to use. Moreover, as artificial scientific text can be generated by different language models, it is difficult for experts to identify which text might originate from the same source beforehand, making it impractical to apply the most appropriate model to each one of them. Therefore, our workflow should facilitate effective and efficient detection of multiple scientific text.

\section{Workflow \& System Design}
\label{sec:workflow}

We present a mixed-initiative workflow (Fig.~\ref{fig:workflow}) that integrates human experts' prior knowledge with multiple ML models for artificial scientific text detection to fulfill the design requirements. In this section, we first describe the four stages of our proposed workflow. Then, we introduce the visual designs and interactions of our prototype system (Fig.~\ref{fig:teaser}) based on the workflow.
% Finally, we provide a concise overview of our implementation details, including dataset collection, model training, and system development.

\subsection{Workflow}
\label{sec:workflow_desc}

\subsubsection{Human Evaluation}

In the first stage, users are presented with scientific text excerpts to review (Fig.~\ref{fig:workflow}$A$). At the entry point of the workflow, these excerpts are either manually selected by the users based on their familiarity with the topic, or recommended randomly by the system. However, in subsequent iterations of the workflow, the selection of excerpts is automated based on the weighted similarity of feature distribution and contribution (Sec.~\ref{subsec:similarity}). We ask users to represent their prior knowledge as \emph{weak annotations} of the text excerpts (R1). While previous literature and our formative study have demonstrated the ability of human experts and ML models to distinguish between machine-generated and human-written scientific text, we believe that they are complementary. These weak annotations can be supported and refined by ML models in subsequent stages of the workflow. Additionally, we employ various visualizations such as area charts and highlightings to enable users to interactively analyze features (R1). Although users may not initially understand the meaning and effects of each feature, we posit that after several iterations, they can gradually acquire knowledge of key features and improve their detection ability by integrating their prior knowledge with feature analysis. Notably, users can also apply the outputs of an ML model as weak annotations in batches if they feel confident about its performance on the current text excerpts to increase efficiency (R4).

\subsubsection{Model Inspection}

In the second stage, we incorporate machine intelligence by comparing human experts' decisions with those made by ML models (Fig.~\ref{fig:workflow}$B$). First, we run all the available ML models on the current text excerpts to obtain their classification results (R2). These ML models are trained using the statistical features derived from our formative study to make their behavior and performance more understandable to human experts, as these features are closely related to the \emph{critical distinctions} identified by human experts. However, different ML models may be trained on different datasets, resulting in varying performances when detecting machine-generated text from different origins. For instance, a model trained on text generated by GPT-3 may not accurately detect text generated by ChatGPT. As human experts can holistically identify machine-generated scientific text, we rank the ML models based on their \emph{level of prior knowledge matching} to human experts, which is computed as

\vspace{-3mm}
\begin{equation}
\label{eq:matchscore}
LevelofPKMatching = (1-\omega_g) * R_{local} + \omega_g * R_{global}
\end{equation}

\noindent
where $R_{local}$ and $R_{global}$ indicate the rate which the classification results of each model \emph{match} those of the user in the \emph{current iteration} and \emph{all previous iterations}, respectively. $R_{global}$ is weighted by a parameter $\omega_g$, which can be adjusted by the user at any point of the workflow according to their preferences and requirements (R1). 
% A lower $\omega_g$ means that the user focuses more on the current iteration and hopes to always find the most appropriate model for the current iteration. For instance, suppose that the machine-generated text in current iteration all come from GPT-3, the user wants to find the exact model that performs the best on detecting text generated by GPT-3. Constrastively, a higher $\omega_g$ means that the user focuses more on the entire detection process without paying too much attention on local minimums. That is, to find the model that performs moderately on detecting text generated by different LLMs.
Lower values of $\omega_g$ indicate a focus on finding the best model for the current iteration, while higher values indicate a broader focus on the entire detection process. Notably, We rank ML models not only to meet users' needs for selecting the most appropriate model (R2), but also to form \emph{positive first impressions} which make it more likely for users with domain expertise to trust and use the intelligent system in the future, as suggested in \cite{Nourani_King_Ragan_2020}. Finally, we use SHAP values\cite{NIPS2017_8a20a862} to generate contribution-based explanations of statistical features to facilitate interpretability (R3).

\subsubsection{Human Confirmation}

In the third stage, users can review the list of ML models ranked by \emph{level of prior knowledge matching} (Fig.~\ref{fig:workflow}$C$). The most appropriate model for the current iteration (either globally or locally) should ideally be ranked high. Users can validate their weak annotations by investigating the feature distributions and contributions (R1). While the top-ranked model is initially selected for analysis, users are free to explore any other models to gain insights into why a particular model performs poorly or to validate their prior knowledge. If all models fail to match users' prior knowledge well, this may be due to inadequate training of the models or lack of expertise of the users. In such cases, users may refine their prior knowledge and assess the usability of the models by analyzing the low-ranked models. Subsequently in this stage, users are responsible for resolving any conflicts arising from different models' decisions and confirming the final results for the text excerpts (R1, R2). We provide effective visualizations to assist users in identifying models' decisions that contradict their weak annotations, and to enable model-wise comparisons for each text excerpt (R1).

\subsubsection{Text Recommendation}
\label{subsec:similarity}

In the fourth stage, we perform similarity calculation to recommend text excerpts for the next iteration of the workflow (Fig.~\ref{fig:workflow}$D$). We calculate the average cosine similarity between each remaining text excerpt and the current text excerpts as follows

\vspace{-3mm}
\begin{equation}
\label{eq:sim}
sim_i = \frac{1}{n} \sum_{j=1}^{n} \Bigg( \omega_d \cdot \frac{f_{d,j} \cdot f_{d,i}}{\|f_{d,j}\|\|f_{d,i}\|} + (1-\omega_d) \cdot \frac{f_{c,j} \cdot f_{c,i}}{\|f_{c,j}\|\|f_{c,i}\|} \Bigg)
\end{equation}

\noindent
where $n$ indicates the iteration size, and $f_{d,i}$ and $f_{c,i}$ indicate the feature distribution and contribution vector of the $i$-th text excerpt, respectively. The similarity of feature distribution is weighted by a parameter $\omega_d$. Higher values of $\omega_d$ indicate a preference for handling text excerpts generated by similar LLMs first, as they typically have similar feature distributions. Conversely, lower values of $\omega_d$ indicate a preference for handling text excerpts treated by similar \emph{model strategies} first. As suggested by Collaris\etal\cite{collaris2022strategyatlas}, model strategies reflect different treatments of ML models to the input data. In the context of scientific text detection, we formulate model strategies as various types of \emph{characteristics} of text. For instance, some text excerpts may be classified as machine-generated mainly due to their short paragraph length, while others may be due to their abnormal writing styles. Users are allowed to adjust $\omega_d$ freely at any point of the workflow to balance the handling of text excerpts from similar origins and those with similar characteristics (R1). As contribution-based explanations are model-specific, we use the feature contributions of the top-ranked model in the current iteration to calculate the weighted similarity. We then select the text excerpts with the highest similarities among the remaining ones for the next iteration. Iteratively presenting several text excerpts enhances the effectiveness and efficiency of detecting multiple text excerpts (R4), since users can conveniently adopt the most appropriate model of the current iteration to the next one without much performance loss. Additionally, if users are unsatisfied with the recommended text excerpts, they can reduce the iteration size to achieve a more fine-grained detection process.

The fourth stage marks the end of an entire iteration, and the workflow continues to loop from the first stage until all text excerpts are confirmed by the user. Through an iterative manner, the workflow facilitates the effectiveness and efficiency in detecting multi-sourced scientific text, and provides sufficient support for human-machine collaboration and interpretation of ML models' decisions.

\subsection{User Interface}

\subsubsection{Text Overview}
\label{sec:text_overview}

The \emph{Text Overview} (Fig.~\ref{fig:teaser}$A$) panel provides an interactive tabular layout that shows the text excerpts in the current iteration of the workflow. It allows users to compare and analyze decisions made by themselves and different models either in a model-wise (vertically) or instance-wise (horizontally) manner. Each text excerpt is encoded as a circle with different colors representing their labels, and the predicted probability of each model is shown as the size of the fan-shaped area inside each circle. Models are ranked from left to right based on their \emph{level of prior knowledge matching} in the current iteration, and users can click on each column to focus on the corresponding model. Besides, clicking on each row allows users to analyze and annotate the corresponding text excerpt. The current focused text excerpt or model is highlighted in light grey. To the right of each model's name is a small barchart (\icon{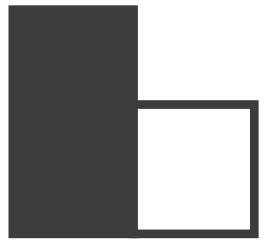}) indicating the \emph{Local Match Rate} and \emph{Global Match Rate}. Additionally, users can apply the outputs of a model as weak annotations by clicking on the \emph{`batch apply'} icon (\icon{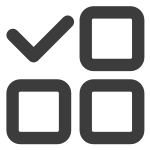}) to the left of each model's name. Above the tabular layout are the number of labeled text excerpts, current iteration status, and model ranks, and below are sliders for adjusting the workflow parameters (Sec.~\ref{sec:workflow_desc}) and a button for submitting decisions. Notably, changes to the parameters will take effect in the next iteration.

This panel serves different purposes depending on the stage of the workflow. In the first stage, the circles representing models' decisions are not displayed (unless users wish to \emph{`batch apply`} the outputs of a particular model), as users need to provide weak annotations based on their prior knowledge. In subsequent stages, all results of models are displayed, as users need to compare and confirm the final results for the current text excerpts.

% \begin{figure}[t]
%     \centering
%     \includegraphics[width=\linewidth]{figs/feature.png}
%     \caption{Distribution-level analysis of (A) numerical, (B) low-dimensional, and (C) high-dimensional features. Relevant text excerpts for comparison can be selected through brushing in ($A_1, B_1, C_1$).}
%     \label{fig:feature}
%     \vspace*{-4mm}
% \end{figure}

\subsubsection{Feature Exploration}

The \emph{Feature Exploration} (Fig.~\ref{fig:teaser}$B$) panel allows users to gain an overview of the feature distribution and contribution of all text excerpts to be detected. It consists of two UMAP\cite{mcinnes2018umap} projections: one for feature values and the other for feature contribution values. Text excerpts are represented as colored dots according to their corresponding labels, with unlabeled excerpts being depicted in grey. Those in the current iteration are highlighted in red, and users can manually select other excerpts to review by brushing. Clusters in \emph{feature distribution} typically correspond to text generated by similar LLMs, while clusters in \emph{feature contribution} indicate text treated by similar model strategies. Notably, clusters in the two projections generally do not match with each other as there is usually no explicit relationship between feature values and feature contribution values. By examining the 2D projections of feature distribution and contribution, users can quickly identify patterns related to different language models or model strategies, as well as the relationship between text excerpts across consecutive iterations. For instance, if the text excerpts in the current iteration are similar to those in the previous iteration in terms of feature distribution, users may choose to \emph{`batch apply'} the outputs of the previous best model to improve efficiency.

\subsubsection{Text Analysis}

The \emph{Text Analysis} (Fig.~\ref{fig:teaser}$C$) panel provides three levels of feature analysis for the selected text excerpt in the \emph{Text Overview} panel.
% Our statistical features are derived from the \emph{critical distinctions} identified through a quantitative experiment and are grouped into three major dimensions (\ie \emph{syntax}, \emph{semantics}, and \emph{pragmatics}) and several subcategories, as discussed in Sec.~\ref{sec:formative_study}.

\textbf{Contribution-level analysis.} As shown in Fig.~\ref{fig:teaser}$C_2$, we display each distinction category in a grouped style similar to tab groups in web browsers. This facilitates easy comparison and analysis of the impact of different features. Each feature dimension is presented as a separate \emph{tab} using different colors, and comprises several subcategories represented by \emph{sub-tabs}, which are outlined in the same color as their corresponding dimension for clarity. Users can sort the feature dimensions and subcategories on the tab strip, and can expand or collapse a feature dimension by clicking on its tab. Clicking on a subcategory tab highlights it and allows users to view the associated feature details presented as multiple \emph{feature cards} (Fig.~\ref{fig:teaser}$C_3$).

We calculate the aggregate contribution values of each feature dimension or subcategory by summing up the SHAP values of the included features\cite{9555810, shapley1953value}. The contribution values for each feature dimension and subcategory are depicted as a vertical bar below each corresponding tab (Fig.~\ref{fig:teaser}$C_2$), with lighter colors indicating subcategories and darker colors representing dimensions. For each included feature, its contribution value is shown as a horizontal bar atop the corresponding feature card (Fig.~\ref{fig:teaser}$C_3$). Notably, negative contribution values are indicated by striped bars. By presenting contribution values in a coarse-to-fine-grained manner, we enable users to analyze features at different levels of granularities catering to their analysis goals. For example, some users may only be interested in understanding the most critical distinctions affecting the models' decisions, while others may wish to delve deeper to gain a more comprehensive understanding of the underlying statistical features' effects.

\textbf{Distribution-level analysis.} It may be insufficient to only display the numerical feature values and their contribution values for users to comprehend and build trust in models' decisions, as some statistical features may be unfamiliar to them (\eg \emph{Gunning-Fog Index}). To this end, inspired by \cite{9555810}, we adopt \emph{reference values}, which are calculated from a relevant cohort of text excerpts. We obtain the 500 most similar text excerpts to the current selected one from the focused model's training dataset based on cosine similarities between their feature distribution vectors. Users can also manually select a more fine-grained cohort by brushing in the UMAP projection of the default relevant excerpts in \emph{Cohort Selection} (Fig.~\ref{fig:teaser}$C_3left$).

The statistical features include numerical values and multidimensional vectors, which are displayed in \emph{Feature Comparison} (Fig.~\ref{fig:teaser}$C_3right$). We employ area charts to visualize the distributions of numerical features (\eg \emph{paragraph length}) and violin charts to visualize the distributions of vector elements for low-dimensional features that consist of meaningful elements (\eg \emph{part-of-speech tag frequency}). For high-dimensional features, such as embeddings from a transformer model, we leverage scatterplots to visualize the UMAP projection of the feature vectors. The same color encodings as before are used to depict machine-generated and human-written text. Moreover, we indicate feature values of the current selected text excerpt with a vertical line in area charts, a horizontal line in violin charts, and an arrow in scatterplots. Reference values are determined by lower and upper bounds (for numerical features) or clusters (for embedding features) of the relevant cohort. By comparing the current feature value with its corresponding reference values in context, users can determine whether the text excerpt is abnormal in terms of each feature. For instance, if the \emph{Gunning-Fog Index} value of the current text excerpt is significantly lower than those of similar human-written text, it may indicate that the text is likely machine-generated.

\textbf{Excerpt-level analysis.} Some features are calculated based on word frequencies in the text excerpt, such as \emph{part-of-speech tag frequency}. Users prefer to understand these features in the context of the original text. Therefore, we visually associate these features with the corresponding raw text. As shown in Fig.~\ref{fig:teaser}$C_1$, by clicking on a feature element in the violin charts, users can highlight the associated words in the raw text. This allows users to gain a deeper understanding of the correlations between statistical features and specific characteristics in the original text. By analyzing back and forth between features and text, users can gradually learn from the model's decisions and explanations and become more attentive to those characteristics in future detection processes. However, some features, such as \emph{average sentence length}, are relative to the entire text rather than individual words. Therefore, we do not provide any interactions for reference, and users should observe their distributions as a whole.

Notably, feature contribution values are not displayed in the first stage of the workflow, whereas in subsequent stages, feature distribution and contribution values are subject to the current focused model.
\begin{figure*}[ht]
    \centering
    \includegraphics[width=\textwidth]{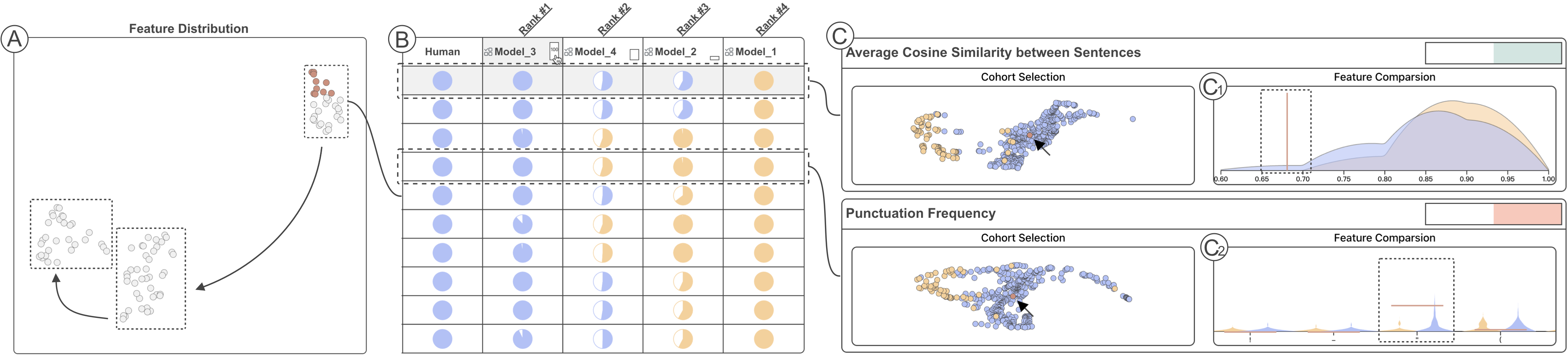}
    \caption{Case Study 1 illustrates how the user (A) locates clusters in \emph{feature distribution}, (B) reviews models' decisions, and (C) builds trust in the appropriate model through feature analysis.}
    \label{fig:case1}
    \vspace*{-3mm}
\end{figure*}

\section{Evaluation}

We trained 4 models to distinguish machine-generated and human-written scientific text on the datasets described in Sec.~\ref{sec:formative_study}. Table~\ref{tab:model_results} shows that the accuracy scores of the models are high when trained and evaluated on the same dataset, but their performance decreases across different datasets due to the OOD issue mentioned earlier. We also trained a model on a combined dataset, which performed relatively well but not optimally. Based on these models and our prototype system, we conducted two case studies and a controlled user study to demonstrate the effectiveness of our approach.

\subsection{Case Study 1: Origin-based Detection}

P2 aimed to review 100 manuscripts and identify which ones were likely to be machine-generated. With over 10 years of research experience, P2 felt confident in his ability to detect machine-generated scientific text. He hoped to improve his efficiency in detecting multiple text excerpts and to seek evidence from ML models to support his own decisions.

\textbf{Explore feature distribution and contribution.} P2 began by checking the \emph{Feature Exploration} panel and observed three clusters in \emph{feature distribution} (Fig.~\ref{fig:case1}$A$), indicating that the corresponding text excerpts likely shared similar origins. P2 recognized that each cluster of text excerpts could be detected using the same ML model for optimal performance. He then decided to adopt an \emph{origin-based} procedure and set \emph{Global Match Weight} to $0.1$ to direct the system to present the locally optimal model in each iteration. This allowed him to focus solely on the top-ranked models and avoid analyzing all the results tediously.

\textbf{Provide weak annotations.} At the start of the workflow, the system automatically presented the text excerpts for the first iteration, using the default iteration size of $10$. All of these excerpts were from the cluster located in the upper right corner of \emph{feature distribution}. P2 provided weak annotations for each text excerpt based on his prior knowledge. However, he did not extensively analyze the features, as they were not familiar or intuitive to him at the moment. P2 hoped to delve deeper into these features once the models' decisions were displayed.

\textbf{Review models' decisions.} Upon submission of the weak annotations, all four models were run. As shown in Fig.~\ref{fig:case1}$B$, \emph{Model\_3} was the top-ranked model with a \emph{local match rate} of $100\%$. P2 then selected each text excerpt to investigate their feature distribution and contribution values. The significant contributions were mainly made by two subcategories: \emph{Coherence} and \emph{Grammatical Issues}. In the \emph{Coherence} subcategory, the most important feature was \emph{Average Cosine Similiarity between Sentences} whose value was obviously lower than the corresponding distribution of human-written text (Fig.~\ref{fig:case1}$C_1$). This finding was consistent with P2's own evaluation, who commented that the paragraph was not logically connected. In the \emph{Grammatical Issues} subcategory, the primary feature was \emph{Punctuation Frequency}, which is a multidimensional feature displayed using a violin chart. P2 discovered that the frequencies of \emph{colons} and \emph{quotation marks} were far beyond the reference value range of human-written scientific text (Fig.~\ref{fig:case1}$C_2$), indicating that the text was likely machine-generated. P2 confirmed this observation, noting that scientific abstracts generally do not contain such punctuation.

\textbf{Re-adjust similarity calculation strategy.} After completing the first iteration, the system presented another $10$ text excerpts for the second iteration. However, P2 noticed that they were not in the same cluster as before. This was due to the default value of \emph{Feature Distribution Weight} being set to $0.5$, which did not prioritize similar text excerpts. Therefore, P2 manually re-selected the text excerpts and adjusted the parameter to $0.9$ to ensure the system recommended text excerpts from similar origins.

\textbf{Batch apply models' decisions.} P2 chose to `batch apply' the outputs of \emph{Model\_3} to the current text excerpts due to their high similarities with the previous ones. However, he still manually verified the model's decisions for each excerpt to ensure consistency with his prior knowledge. Notably, only one text excerpt failed to match due to some semantic errors that were difficult for models to capture but obvious to human experts. After careful analysis, P2 corrected the model's decision for this case. Through the whole process, P2 gradually built trust in \emph{Model\_3}.

\textbf{Move on to other clusters.} P2 proceeded to examine the text excerpts in the second cluster located in the lower left corner. This time, \emph{Model\_2} became the top-ranked one as it performed well on the current text excerpts that were possibly from another origin. Therefore, P2 followed similar procedures as before by extensively analyzing only the initial text excerpts and then `batch applying' the model's decisions to those in the same cluster. This enabled him to efficiently complete the detection process for all clusters in \emph{feature distribution}.

\textbf{Summary.} This case study shows how our approach assists human experts in efficiently detecting multi-sourced text by iteratively handling text excerpts from similar origins and utilizing ML models to provide evidence to support their judgments through feature analysis.

\begin{table}
\renewcommand\arraystretch{1.1}
\small
\centering
\caption{Accuracy scores of the ML models trained and evaluated on datasets of different LLMs.}
\label{tab:model_results}
\setlength{\tabcolsep}{4.5pt}
\begin{tabular}{c|cc|cc|cc|cc}
\toprule
\multirow{3}{*}{\bf Training Data} & \multicolumn{8}{c}{\bf Test Data} \\ \cline{2-9} 
                               & \multicolumn{2}{c|}{ChatGPT} & \multicolumn{2}{c|}{GPT-3} & \multicolumn{2}{c|}{GPT-2} & \multicolumn{2}{c}{Total} \\ \cline{2-9} 
                               & Acc. & AUC & Acc. & AUC & Acc. & AUC & Acc. & AUC \\ \hline
ChatGPT                        & 0.953    & 0.953   & 0.891    & 0.891   & 0.576    & 0.589   & 0.794    & 0.807   \\ \hline
GPT-3                          & 0.783    & 0.796   & 0.975    & 0.975   & 0.443    & 0.393   & 0.697    & 0.714   \\ \hline
GPT-2                          & 0.530    & 0.697   & 0.473    & 0.381   & 0.988    & 0.987   & 0.692    & 0.794   \\ \hline
Total                            & 0.816    & 0.817   & 0.908    & 0.914   & 0.865    & 0.869   & 0.856    & 0.859   \\ \bottomrule
\end{tabular}
\vspace{-2mm}
\end{table}

\subsection{Case Study 2: Strategy-based Detection}

P6 was provided with 40 manuscripts to review. Unlike P2, P6 was not as experienced and hoped to benefit more from human-machine collaboration and learn from models' decisions through comprehensive feature analysis.

\begin{figure*}[t]
    \centering
    \includegraphics[width=\textwidth]{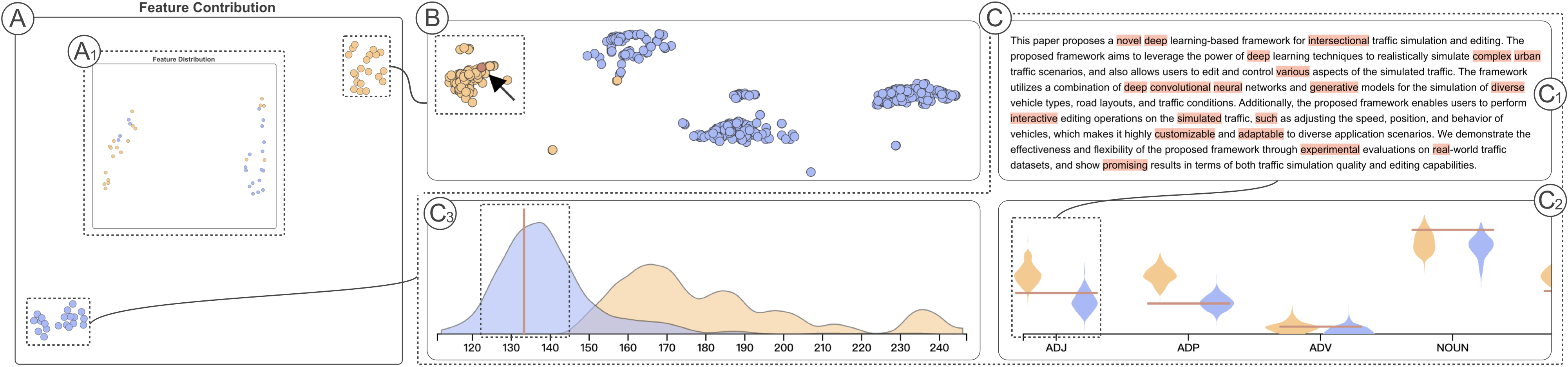}
    \caption{Case Study 2 illustrates how the user (A) locates clusters in \emph{feature contribution}, (B) analyzes the human-written text cluster, and (C) analyzes the machine-generated text cluster.}
    \label{fig:case2}
    \vspace*{-4mm}
\end{figure*}

\textbf{Locate similar model strategies.} P6 examined the \emph{Feature Exploration} panel and identified two clusters in \emph{feature contribution} (Fig.~\ref{fig:case2}$A$), indicating similar model strategies. As such, P6 decided to adopt a \emph{strategy-based} procedure. He adjusted \emph{Global Match Weight} to $0.9$ to avoid local optimum and set \emph{Feature Distribution Weight} to $0.1$ to ensure the system recommended text excerpts from the same clusters in \emph{feature contribution}.

\textbf{Find the globally best model.} After several iterations, P6 discovered that only \emph{Model\_4} had a moderate \emph{global match rate} of 80\%. The separate locations of the text excerpts in \emph{feature distribution} (Fig.~\ref{fig:case2}$A_1$) suggested that they came from diverse origins, resulting in poor performances of the other three models trained on single datasets. In contrast, \emph{Model\_4} was trained on a combined dataset and showed an acceptable accuracy in detecting multi-sourced text (Tab.~\ref{tab:model_results}). Hence, P6 decided to conduct an analysis focused on \emph{Model\_4} to gain some insights.

\textbf{Analyze the first cluster.} It turned out that most text excerpts in the cluster located in the lower left corner of \emph{feature contribution} were machine-generated. To investigate the most influential factors, P6 inspected the \emph{Text Analysis} panel for each text excerpt. He found that the \emph{Syntax} dimension consistently contributed the most, with the \emph{Text Structure} subcategory having the greatest impact within this dimension. Furthermore, the contribution value of \emph{Word Count} was much higher than others. P6 then performed a more fine-grained analysis by brushing a smaller area in \emph{Cohort Selection}, and found that the \emph{Word Count} value was significantly lower than most human-written text (Fig.~\ref{fig:case2}$C_3$). Additionally, \emph{Part-of-Speech Tag Frequency} in the \emph{Grammatical Issues} subcategory was another important feature. Further investigations revealed that the frequencies of certain word classes were noticeably lower than those in human-written text (Fig.~\ref{fig:case2}$C_2$). By highlighting the corresponding words in the raw text (Fig.~\ref{fig:case2}$C_1$), P6 gained a deeper understanding about this feature. Notably, these observations persisted across most text excerpts of the first cluster, indicating that they were classified as machine-generated primarily due to syntax issues.

\textbf{Analyze the second cluster.} On the other hand, most text excerpts in the cluster located in the upper right corner of \emph{feature contribution} were human-written. Further investigations revealed that the \emph{Pragmatics} dimension always had the highest positive contribution, whereas the other two dimensions had negative or negligible contributions. Therefore, P6 analyzed the most influential feature \emph{SciBert Embedding}, which belonged to the \emph{Writing Style} subcategory. By comparing the current focused text excerpt (red dot) with the relevant cohort, P6 discovered that it belonged to a cluster predominantly comprising human-written text (orange dots), while other clusters were mainly composed of machine-generated text (blue dots), as shown in Fig.~\ref{fig:case2}$B$. This observation was also validated in other text excerpts of the second cluster, suggesting that they were classified as human-written mostly based on their writing styles.

\textbf{Gain knowledge from models.} Through the analysis of different model strategies, P6 outlined two characteristics that may help him in future identification of scientific text. First, machine-generated scientific text are typically shorter and less detailed compared to human-written ones. Second, the writing styles of human-written scientific text differed significantly from machine-generated ones in terms of specificity and formality. These principles align with our formative study and are further supported through feature analysis.

\textbf{Summary.} This case study demonstrates how our approach assists human experts in interpreting models' decisions by iteratively handling text excerpts treated by similar model strategies and leveraging models' insights to identify distinctive characteristics that differentiate machine-generated text from human-written text.

\subsection{User Study}

\subsubsection{Experiment Design}

\textbf{Participants.} We collaborated with the 12 experts participated in the formative study, who had sufficient domain expertise and were familiar with the workflow and prototype system.

\textbf{Settings.} To evaluate the effectiveness and efficiency of our approach, a comparative study was conducted using three conditions for scientific text detection:

\emph{C1. Human only.} An ablated version of the prototype system was utilized that only displayed raw text. Participants annotated each text excerpt based on their prior knowledge.

\emph{C2. Machine only.} An ablated version of the prototype system was utilized that only displayed the outputs of four models. Participants applied one of the outputs to each text excerpt.

\emph{C3. Human-machine collaboration.} The complete version of the prototype system was utilized. Participants examined each text excerpt combined with feature analysis and then confirmed their final decisions.

\begin{figure}
    \centering
    \includegraphics[width=\linewidth]{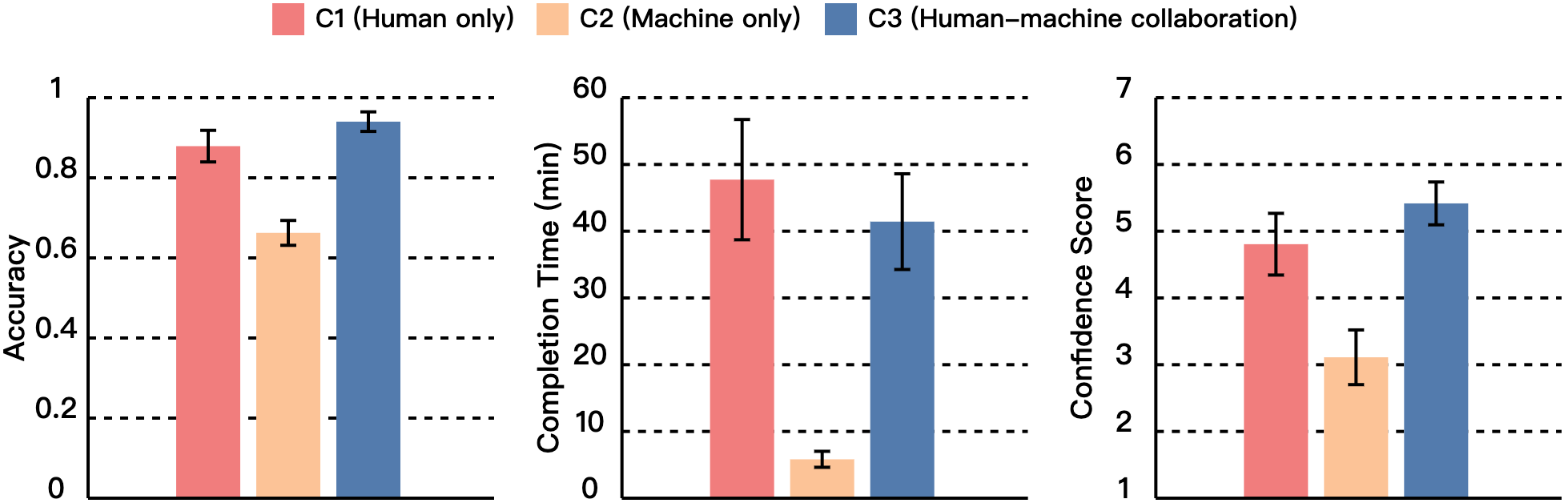}
    \caption{Overall accuracy, total completion time, and average confidence score for each condition.}
    \label{fig:user_study}
    \vspace*{-2mm}
\end{figure}

\textbf{Procedure.} Participants were introduced to the study's purpose and asked to complete a consent form. They were then instructed on the experiment design and encouraged to familiarize themselves with the three versions of the prototype system. During the experiment, the order of conditions were counterbalanced across participants. For each condition, we sampled 60 text excerpts (15 ChatGPT-generated, 15 GPT-3-generated, 15 GPT-2-generated, and 15 human-written), yielding 180 trials in total. In addition to annotations, participants were required to rate their \emph{confidence score} on a 7-point Likert scale from 1 (\emph{low confidence}) to 7 (\emph{high confidence}) for every 10 trials\cite{Nourani_King_Ragan_2020}. Finally, we evaluated the results for each condition in terms of \emph{effectiveness}, \emph{efficiency}, and \emph{reliability}, measured by overall accuracy, total completion time, and average confidence score, respectively (Fig.~\ref{fig:user_study}). Post-study interviews were also conducted to collect qualitative feedback.

\subsubsection{Results Analysis}

\textbf{Effectiveness.} \emph{C3} achieved a higher accuracy than \emph{C1} by 7\% on average, both of which outperformed \emph{C2} significantly. This confirms that human experts can detect most machine-generated scientific text, while blindly applying models' outputs proves inadequate in identifying multi-sourced text. Most participants (10/12) increased their accuracy in \emph{C3} compared to \emph{C1}, since the system enabled them to locate the most appropriate model under various conditions by matching their decisions with models' outputs. Participants could then refine their weak annotations based on models' decisions, as certain excerpts were \emph{`initially hard to distinguish'} (P3) or \emph{`required further verification from suitable models'} (P11).

\textbf{Efficiency.} Regarding completion time, \emph{C2} was the fastest by directly utilizing models' decisions but sacrificed accuracy, as most participants resorted `\emph{majority voting}' (P4) or `\emph{random selection}' (P7). Although \emph{C3} was only 13\% faster than \emph{C1} on average, most time was spent on feature analysis or due to \emph{`network lagging'} (P5). Also, our experiment involved a relatively small number of text excerpts, which reduced the gap between completion times. Additionally, participants praised the \emph{`batch apply'} function, which saved much time since they \emph{`do not need to manually check each excerpt anymore'} (P3). Therefore, the efficiency could be further improved once the system issues are resolved, especially for large amounts of text excerpts.

\textbf{Reliability.} The average confidence score in \emph{C3} was 13\% and 74\% higher than \emph{C1} and \emph{C2}, respectively, demonstrating the effectiveness of our approach in improving detection reliability. Compared to \emph{C2}, \emph{C3} significantly increased user trust by allowing them to \emph{`see what factors contributed to models' decisions'} (P7). However, some participants' confidence scores did not increase much between \emph{C3} and \emph{C1} due to explanations being \emph{`beyond expectations'} (P9), potentially caused by some correlated features that made SHAP values less effective\cite{AAS2021103502}. Interestingly, participants tended to seek explanations that supported their initial judgments to \emph{`feel more confident'} (P2), even if the corresponding contribution values were low. This could lead to bias by ignoring other important features. Future research on human-centered XAI\cite{10.1145/3491101.3503727} is needed to explore this phenomenon further.

\textbf{Summary.} Overall, the results showed that our approach facilitated the effectiveness, efficiency, and reliability of the detection process. Despite minor issues such as system lagging, participants gave positive feedback on the usability of our prototype system, especially the feature projections which helped them \emph{`discover interesting patterns in feature distribution and contribution values'} (P4).
\section{Discussion}

In this section, we summarize the implications derived from our work, and then discuss the limitations and future work.

\subsection{Implications}

\textbf{Incorporate human experts' prior knowledge.} Combining the capabilities of human experts and machine intelligence in detecting scientific text is effective, as our work has shown. Human-machine collaboration in the detection process has been previously advocated for\cite{crothers2022machine, jawahar2020automatic, solaiman2019release}, and some tools\cite{gehrmann2019gltr} have been proposed to incorporate a human analyst to facilitate detection. In our work, we match ML models' decisions with human experts' prior knowledge to handle situations where models' outputs conflict with each other or human judgment. We place much emphasis on the human side, as we believe that it is the responsibility of human experts to confirm final judgments in high-stakes decision-making scenarios. Thus, in determining whether a manuscript is fake and violates academic integrity, experts should aim for a balance between trusting their own judgments and seeking evidence from models, rather than blindly accepting models' decisions, especially when models' capabilities may be weaker than humans themselves due to various limitations such as OOD issues.

\textbf{Leverage multiple models to facilitate detection.} Given the diversity and rapid development of LLMs, a single detector may not suffice for all situations. To address the issue, our mixed-initiative workflow is model-agnostic and integrates various detectors to maximize their strengths under different conditions. We assume that human judgments are the `gold standard' in certain scenarios that require domain expertise or context information to verify the detection results, such as academic and educational contexts. Therefore, we provide a new perspective in this work by leveraging multiple detectors along with human agency to enhance the effectiveness and efficiency of the detection process.

\textbf{Support feature analysis with multiple granularities and levels.} Although preliminary works\cite{kowalczyk2022detecting, ma2023abstract} have explored the feature contributions of ML models for detection, such explanations are insufficient for effective interpretation. Our experts appreciated the coarse-to-fine-grained manner to display contribution values that can serve various analysis goals. In addition to contribution-level analysis, we also provide distribution- and excerpt-level analysis to understand features from the cohort and instance perspective, respectively. These analyses enhance human experts' trust in models' decisions and enable them to learn from models' behavior to improve their own detection ability.

\subsection{Limitations \& Future Work}

\textbf{Scalability.} The current prototype system is limited to a maximum iteration size of 30 and 4 ML models in the workflow, which hinders its application in real-world scenarios. Experts noted that the tabular view may not facilitate effective comparisons when there were too many rows or columns, while excessive dots in feature projections may cause visual clutter. Therefore, it is necessary to design more effective visualizations to support the workflow at a larger scale.

\textbf{Generalizability.} The proposed workflow can be extended to various text detection contexts that require human involvement and interpretability, such as identifying cheating in educational contexts. However, given the various \emph{critical distinctions} between machine-generated and human-written text across different scenarios, directly applying the current statistical features and system designs may not be feasible. Additionally, transformer-based models are not compatible with our feature-based explanations and require alternative visualization techniques for interpretation. In future work, we plan to extend the workflow to support newly proposed detection methods such as watermarking\cite{kirchenbauer2023watermark} and probability curvature\cite{mitchell2023detectgpt}.
\section{Conclusion}

In this work, we identify critical distinctions between machine-generated and human-written scientific text through a quantitative experiment. Our findings provide valuable insights into the capabilities of LLMs' in academic writing and can inform the design of more effective detection methods. We propose a mixed-initiative workflow and a visual analytics prototype that incorporates human experts' prior knowledge to facilitate the efficiency, interpretability, and reliability of the detection process. We demonstrate the effectiveness of our approach through two case studies and a controlled user study. We believe that our work will inspire future research on integrating human intelligence into artificial text detection.
\clearpage

\bibliographystyle{ref/abbrv-doi-hyperref}

\bibliography{main}

\end{document}